# EXCHANGE INTERACTION EFFECTS ON THE CONDUCTANCE OF QUANTUM WIRES

Xavier Oriols

*Departament d'Enginyeria Electrònica i Materials*

*Escola Tècnica Superior d'Enginyeria*

*Universitat Autònoma de Barcelona 08193 Bellaterra SPAIN*

Abstract

The effect of exchange interaction on the two-terminal conductance of fully ballistic samples is studied using a many-particle wave packet formalism. The approach shows that the puzzling nonuniversal conductance quantization can be explained as a fermion exchange interaction effect. The Landauer conductance step $G_o \equiv 2e^2/h$ is obtained when two-fermion interaction is considered. Reductions of $G_o$ are due to a many-particle exchange interaction effect. For ballistic GaAs/AlGaAs quantum wires longer than 1 $\mu m$ , a conductance step reduction is obtained in agreement with experimental results.

*Introduction* - The Landauer approach predicts that the two-terminal conductance of a phase coherent sample with ideal transmission is quantized in unit of $G_o = 2e^2/h$ [1]. Experimentally, such conductance quantization was first successfully observed on ballistic quantum point contacts defined in a GaAs-AlGaAs heterostructure verifying the value of the Landauer resistance and the true nature of quantum modes in phase-coherent semiconductors [2,3]. Recently, Yacoby *et al.* [4] have succeeded to produce quantum wires (QW) as long as 1-10 $\mu m$ of extreme quality. Quite surprising, the conductance steps for these samples have constants heights $G_o \cdot \nu$ with $0.5 < \nu \leq 1$ instead of $\nu = 1$ as predicted by the Landauer approach. Such conductance reduction has also been observed by Tarucha *et al.* [5] in similar long QWs. In both sets of experimental results, the conductance steps tends to $G_o$ as the length of the QW is shortened [4,5]. In spite of a great amount of theoretical work [4,6-12] devoted to study these systems, the experimental observations of conductance deviation have found no satisfactory explanation till date.

Basically two paths have been explored in the literature to solve the physics of conductance reduction. The first type of arguments assumes that the coupling between the two dimensional (2D) contacts and the one dimensional (1D) QW is the responsible for such deviations. Elaborated models for the 2D-1D coupling have been proposed [4,6-8] but, although successful in providing the experimental conductance reduction, are not completely satisfactory to explain why the same reduction is observed for each mode. The physics described in these models is based on phenomenological rates such as the scattering from 2D to 1D systems [4,6,7], the backscattering [4,7] or the electron-phonon coupling [8]. As pointed out in ref. 7, one would expect different conductance reduction for each mode due to the energy-dependence of previous rates. The second type of explanations is based on intrinsic properties of a 1D electron gas. In particular, QWs have been studied in the context of Luttinger liquids [9-12]. In the absence of disorder, the conductance of a Luttinger liquids connected to noninteracting leads is given by $G_o$ regardless of the interactions in the QW [9]. On the other hand, it has been shown that the conductance can be reduced due to impurities [10] or disorder [11], but these *dirty* conditions seem not applicable for the ballistic and very clean samples discussed in refs. 4 and 5. Other approaches where charging effects are considered [12] do also suffer from the drawback that the strength of the interaction is determined mainly by the electron density in the QW which makes the explanation of flat plateaus and constant conductance steps difficult. In summary, all previous approaches, although successful for explaining subsets of the experimental results, are not completely satisfactory to resolve the experimental nonuniversal conductance quantization.

In this letter, we will study the role of the fermionic exchange interaction in QWs. We assume a clean QW where each mode is studied as a perfect 1D system. The unavoidable Pauli principle is the only interaction considered [13]. Since the exchange interaction does only affect electrons at the same quantum mode, we can already anticipate that the conductance deviation will be identical for each mode. In particular, we show that the exchange interaction due to one electron is the ultimate origin of the Landauer resistance. On the other hand, we show that the exchange interaction effect between more electrons accounts for a reduction of $G_o$. Our approach provides conductance reductions in good agreement with the experimental results. The effect of the exchange interaction between Fermi particles in 1D systems will be studied within the first quantization framework in the coordinate representation. Fermion interaction will be directly introduced by assuring the antisymmetrical behavior, under the exchange of Fermi carriers, of a many-particle wave function composed of single-particle free wave packets.

*The model* .- We assume a perfectly clean QW of length L gradually opened to the 2D ideal reservoirs, guaranteeing a perfect *adiabatic* connection. Due to the finite cross-section of the QW leading to discrete transverse energy, several conducting channels (i.e. quantum modes) appear. Hereafter, we will study the electron transport in one of these modes considering that it can be perfectly described by a 1D system. We assumed that the perfect reservoirs are characterized by two *outside* chemical potentials $\mu_{L/R}$ (left/right), and that the applied bias, V, is equal to $\mu_L - \mu_R = e \cdot V$ where $e$ is the electron charge. According to the extension of the Ramo-Schockley theorem to semiconductor devices [14], the total instantaneous current, I, is related to the electron motion by:

$$I = \frac{e}{L} \sum_{i=1}^{N} v_i \, , \qquad (1)$$

where N is the number of electrons inside the QW and $v_i$ is the electron velocity. For zero temperature, we define the *active* phase space area available for counting the electrons as $L \cdot (k_L - k_R)$, where the wave vectors $k_{L/R}$ are related to $\mu_{L/R}$ via the energy dispersion relationship, E(k). For DC values, the average electron density will take a constant value, $\rho$, inside the active phase space area and zero elsewhere. In this regard, the DC current obtained from eq. 1 is equal to $<I>= e/L \int_{x=0}^{x=L} dx \int_{k=k_R}^{k_L} dk \cdot \rho \cdot v(k)$ where v(k) is the velocity of the electrons with wave vector k. Using the relation $v(k) = 1/\hbar \cdot dE(k)/dk$, we finally obtain for the DC conductance:

$$G = \frac{e^2}{\hbar}\rho\,. \qquad (2)$$

Although expression 2 is not the usual way of writing the conductance, it specifies a direct link between conductance and electron density in phase space. Let us notice that, by invoking periodic boundary conditions and spin electron degeneracy, the Landauer conductance is obtained (i.e. $\rho = 1/\pi$ ).

The goal of the present work is to directly compute the electron phase space density, $\rho$, for 1D systems. The fundamental point of our approach is that each electron is associated to a quantum state defined by a time dependent wave packet, $\Psi_j(x,t)$. As usual, a linear superposition of Hamiltonian eigenfunctions, $\varphi_k(x)$, is used to describe the wave packet, $\Psi_j(x,t) = \int_{-\infty}^{+\infty} dk \cdot a_j(k) \cdot \varphi_k(x) \cdot e^{-iE(k)\cdot t/\hbar}$, where $a_j(k)$ are the complex time-independent coefficients of such a superposition [15]. For a free Hamiltonian, a normalized free Gaussian wave packet can be defined by:

$$a_j(k) = \frac{1}{(\pi\sigma_K^2)^{1/4}} \exp\left( -\frac{(k-k_j)^2}{2\sigma_k} + i \cdot x_j (k_j - k) \right), \qquad (3)$$

where $x_j$ is the central position and $k_j$ the central wave vector of the wave packet. We define $\sigma_x = 1/\sigma_k$ as the wave packet initial spatial dispersion. As we have noticed previously, the fermionic interaction between the electrons is introduced in the N-particle wave function, constructed from single particle wave-packets, by explicitly taking into account the fundamental Fermion symmetry:

$$\Phi(t;x_1,x_2,....x_N)=\frac{1}{\sqrt{N!}}\sum_{P\in S_N}\left(\prod_{l=1}^{N}\Psi_l(x_{p(l)},t)\right)f(p)$$

Here $S_N$ is the group of the N! permutations p on the set of N electron positions; and f(p) is equal to the sign of the permutations [16]. In this regard, the probability of such a distribution, Q(1,…,N), is defined as $Q(1,...,N)=\int_{-\infty}^{\infty}dx_1.....\int_{-\infty}^{\infty}dx_N\left|\Phi(t;x_1,x_2,....x_N)\right|^2$. For practical proposes it can be computed as:

$$Q(1,...,N)=\sum_{p\in S_N}f(p)\cdot\prod_{l=1}^{N}q(l,p(l)), \tag{4}$$

where we have defined the complex value, $q(l,j)$, as the overlap integral:

$$q(l,j)=\int_{-\infty}^{\infty}a_l^*(k).a_j(k)\cdot dk\,. \tag{5}$$

In particular, for two different Gaussian wave packets described by eq. 3, we obtain:

$$\left|q(l,j)\right|^2=\exp\left(-(k_l-k_j)^2/(2\sigma_k^2)\right)\cdot\exp\left(-(x_l-x_j)/(2\sigma_x^2)\right)=\exp(-d^2), \tag{6}$$

where we have defined the normalized phase-space distance, d, between the central positions of the wave packets as:

$$d^2=(k_l-k_j)^2/(2\sigma_k^2)+(x_l-x_j)^2/(2\sigma_x^2) \tag{7}$$

*Exchange interaction effects due to 1 electron.-* In order to study how the exchange interaction determines the phase-space electron density, let us start by considering two free electrons associated to two different Gaussian wave packets. The central point of one of these wave packets is fixed at $x_1$= 1 $\mu m$ and $k_1$=3 $\mu m^{-1}$ with $\sigma_x = 500nm$. From eqs. 4 and 5, the presence probability of both wave packets is computed as $Q(1,2) = 1 - |q(1,2)|^2$. In fig. 1, we have represented the exchange interaction effect that the first electron will produce on the second one. In particular, we have plotted the probability that the second wave packet is not present due to active presence of the first one, $P(x,k) = 1 - Q(1,2) = |q(1,2)|^2$, where x is the central position and k the central wave vector of the second wave packet. Due to the Pauli principle, the first electron *repels* the second [16]. In this regard, the phase space area not allowed for the second wave packet, A(1), is just a sum over all possible central positions (x,k) weighted by the probability $P(x,k)$:

$$A(1) = \int_{-\infty}^{\infty} dx \int_{-\infty}^{\infty} dk \cdot P(x,k) = \int_{-\infty}^{\infty} dx \int_{-\infty}^{\infty} dk \cdot |q(1,2)|^2 = 2 \cdot \pi . \qquad (8)$$

This result, $A(1) = 2\pi$, is an universal value as far as just one electron inside the active phase space is considered [17]. In average, there is an elliptical area around the first wave packet forbidden for the second one (see inset of fig. 1). Such an ellipse can be defined by the phase space points that accomplish the relationship d=1 in eq. 7 (the area of this ellipse, with axes $\sqrt{8}\sigma_x$ and $\sqrt{8}\sigma_k$, is equal to $A(1) = 2\pi$). Therefore, the maximum phase space electron density is equal to $\rho = 1/\pi$ (a factor 2 for spin) that exactly reproduces Landauer conductance via eq. 2. At this point, let us notice that our approach provides an explanation for the finite value of the conductance (not enough phase space available for additional wave packets) quite close to the standard Fermi liquid argumentation (fully occupation of the available states). In other words, each wave packet of our approach takes the same role as Hamiltonian eigenstates in the standard second quantization framework (Hamiltonian eigenstates are a particular wave packet limit, $\sigma_x \to \infty$). The advantage of our formalism is that we can generalize it to study many particle exchange effects. In particular, we will show that the fermionic interaction due to two or more electrons inside the phase space is the responsible for the conductance reduction.

*Exchange interaction effects due to N electron* - Let us assume a distribution of N wave packets located inside the active phase space with fixed central positions (see contour plots in fig. 2). We are interesting in knowing the probability of such system when another wave packet is added. The quantum probability of the N+1 system can be computed from eqs. 4, 5 and 6. We recognize Q(1,…,N,N+1) as the Slater determinant of the (N+1)x(N+1) matrix, $\{q(l,j)\}$, and it can be easily computed for a large number of particles using text-book standard matrix theory. In fig. 2, we have represented the exchange interaction effects that N=9 wave packets, at fixed positions, will produce on an additional one. As in fig. 1, we have plotted the probability of not finding the system of N+1 wave packets, P(x,k)=1-Q(1,….,N+1), where x and k are the central position and wave vector of the additional wave packet. In fig. 2a, the central positions of the fixed wave packets are chosen to obtain a normalized separation d=4 for all neighboring wave packets (larger than d=2 that defines the average separation of two fermions in previous section). As expected, the N+1 system has a probability equal to one only when the additional wave packet is located far from the elliptical areas surrounding each of the 9 wave packets. On the other hand, in fig. 2.b we have plotted the probability P(x,k) for the same distribution when the normalized distance is equal to d=2. We see that exchange interference effects between all wave packets appears, and the probability of finding a systems with an additional electron has a really low probability independently of its location (in particular, $Q(1,...,9,10) \approx 0.68$ is the highest probability). Clearly, the condition to avoid such a low probability (i.e. such a strong non-local many particle exchange interaction) depends on the separation between wave packets, and for the distribution considered here the separation d=2 is not enough. In order to quantitatively confirm previous results, we compute the dependence of the presence probability Q(1,..,N+1) on the distance d, for several values of N. As depicted in the scheme of fig. 3, now, we assume a set of N+1 wave packets with the same central wave vector, but different central position separated by a constant normalized distance d. The fig. 3 confirms that for identical value of d, we obtain Q(1,..,N)>Q(1,..,N,N+1) as it is clearly depicted in the inset (the horizontal line corresponds to the previously mentioned value $Q(1,2) = 0.982 \quad for \quad d = 2$). The main conclusion is that the separation between wave packets must increase with N in order to provide distribution of wave packets with probabilities close to unity. Let us notice that we have checked this conclusion for other arrangements of electrons (the distribution considered in fig. 2 are depicted in dotted line in fig. 3) confirming this result. The surprising consequence is that the conductance will be reduced since the phase space density will decrease. By defining A(N) as the phase space area occupied by each of the N interacting wave packets and using $\rho(N) = 1/A(N)$, the conductance of a N-particle systems can be written as $G(N)/G(1) = A(1)/A(N)$. We have previously shown that $A(1) = 2\pi$ and $G(1)=G_o$; therefore if A(N) grows (i.e. larger separation d) the conductance G(N) decreases.

In order to quantitatively compare our predicted conductance reduction with the experimental results found in ref. 4, we provide an approximate expression for the distance between wave packets, d(N), as a function of N. From fig. 3, all distributions of wave packets with presence probability close to one have separations larger than d=2. In this range ($d \geq 2$), the computation of Q(1,..,N,N+1) for the distributions depicted in fig. 3 can be greatly simplified by realizing that mainly all overlaps integrals q(l,j) are negligible in comparison to $q(l,l)=1$ and $|q(l,l\pm 1)|^2 = \exp(-d^2)$. Therefore, the probability of such a N distribution of wave packets is just the determinant of the quasi-diagonal matrix $\{q(l,j)\}$ that can be simply written as $Q(1,..,N,N+1) \approx 1 - N \cdot \exp(-d^2)$ (exactly the same expression is obtained for a unique central position and N+1 regular different central wave vectors). Since the role of the reservoir (not considered yet) is just to favour those distributions that transfer the maximum number of particles, we assume that a reasonable distribution takes the probability value Q(1,…,N,N+1)=$P_e$=0.982 (in accordance with $Q(1,2) = 0.982$ for d=2). Therefore, we can compute the distance between wave packets as $d(N)^2 = \ln\left(N/(1-P_e)\right)$. From previous consideration, the phase space area A(N) is defined as an ellipse with axes $\sqrt{8}\sigma_x d(N)/2$ and $\sqrt{8}\sigma_k d(N)/2$, so that $A(N) = \pi \cdot d(N)^2/2$. On the other hand, the number of interacting electrons inside the active phase space is related with the sample length L via $N \cdot A(N) = L \cdot (k_L - k_R)$. Since the applied voltage in ref. 4 is approximately $10\,\mu V$, by assuming an effective mass for AlGaAs of 0.067 times the free electron mass and adding the thermal energy at 0.3 $^o$K, we obtain $(k_L - k_R) \approx 1 \cdot 10^7 m^{-1}$. With these results, in fig. 4 we have represented the conductance quantization values as a function of the length L of a AlGaAs-AsGa QW. The experimental results of ref. 4 are also plotted for fully ballistic samples in good agreement with our results within the experimental uncertainties. In QW shorter than 1 $\mu m$ (at low bias and temperature) only one electron is present in the phase space. The Pauli principle associated to this electron prevents a second electron to enter, limiting the conductance to $G_o$. On the other hand, a reduction of $G_o$ is obtained for longer samples due to the exchange interaction effect associated to many (>1) electrons present in the phase space.

*Conclusion:-* In this letter we have discussed the origin of the puzzling non-universal conductance quantization in QW. We have shown that such quantization is ultimately due to the Pauli Exclusion Principle. A two-fermions system provides a maximum conductance equal to the well-known Landauer value $G_o$ and conductance reductions are obtained for many-fermions systems. On the other hand, since electrons at different quantum modes do not suffer from exchange interaction, all the result discussed in this letter for an ideal 1D systems are identical for each quantum mode, in agreement with experimental results. Finally, apart from enlightening the physical origin of the conductance quantization, the present letter describes a novel approach (that we call the many particle wave packet formulism) that provides a new framework to study quantum transport. Extensions of the present approach to nonzero temperature, Bose carriers or partially transparent barriers will provide a powerful framework to study dynamic properties of systems that deal with few particles as is becoming common in the nowadays very active field of nanoelectronics.

The author is grateful to State University of New York at Stony Brook, where this work was carried out, for its charming hospitality.

[17] The result is independent of any wave packet parameter such as the central position, central wave vector, spatial dispersion or initial time. For example, the wave packet spreading (not explicitly considered) that will tend to extend the overlap integral is exactly compensated by a reduction of the wave function modulus providing $A(1) = 2\pi$ . Just to deal with simpler analytical expression, it is assumed that the dynamics are smaller than the decoupling time as defined in ref. J.Suñé and X. Oriols, Phys. Rev. Lett., vol.85, 894 (2000).

**Figure caption:**

Figure 1: Probability of not finding a second wave packet, due to the presence of the first one, P(x,k), as a function of the central position, x, and wave vector, k, of the second wave packet. The inset shows P(x,k) in a contour plot. The contour line P(x,k)=0.36 corresponds to an ellipse with an area equal to $2\pi$.

Figure 2: Exchange interaction effect due to a distribution of N=9 wave packets with different positions and identical $\sigma_x = 500nm$. P(x,k) is the probability of not finding and additional wave packet as a function of its central position, x, and wave vector, k. (a) d=4. (b) d=2.

Figure 3: In solid lines, the probability Q(1,..,N+1) as a function of d for the distribution depicted in the scheme. In dashed lines, Q(1,..10) for the distribution of fig. 2 when the additional wave packet is located with the same normalized separation. The arrows indicate the distances d=2 (fig 2b) and d=4 (fig2a). The horizontal dashed line in the inset correspond to the probability $Q(1,2) = 0.982$ for d=2.

Figure 4: Conductance reduction, in unit of the Landauer conductance $G_o$, as a function of a GaAs-AlGaAs quantum wire length L. Experimental results from ref. [4] are indicated by circles.

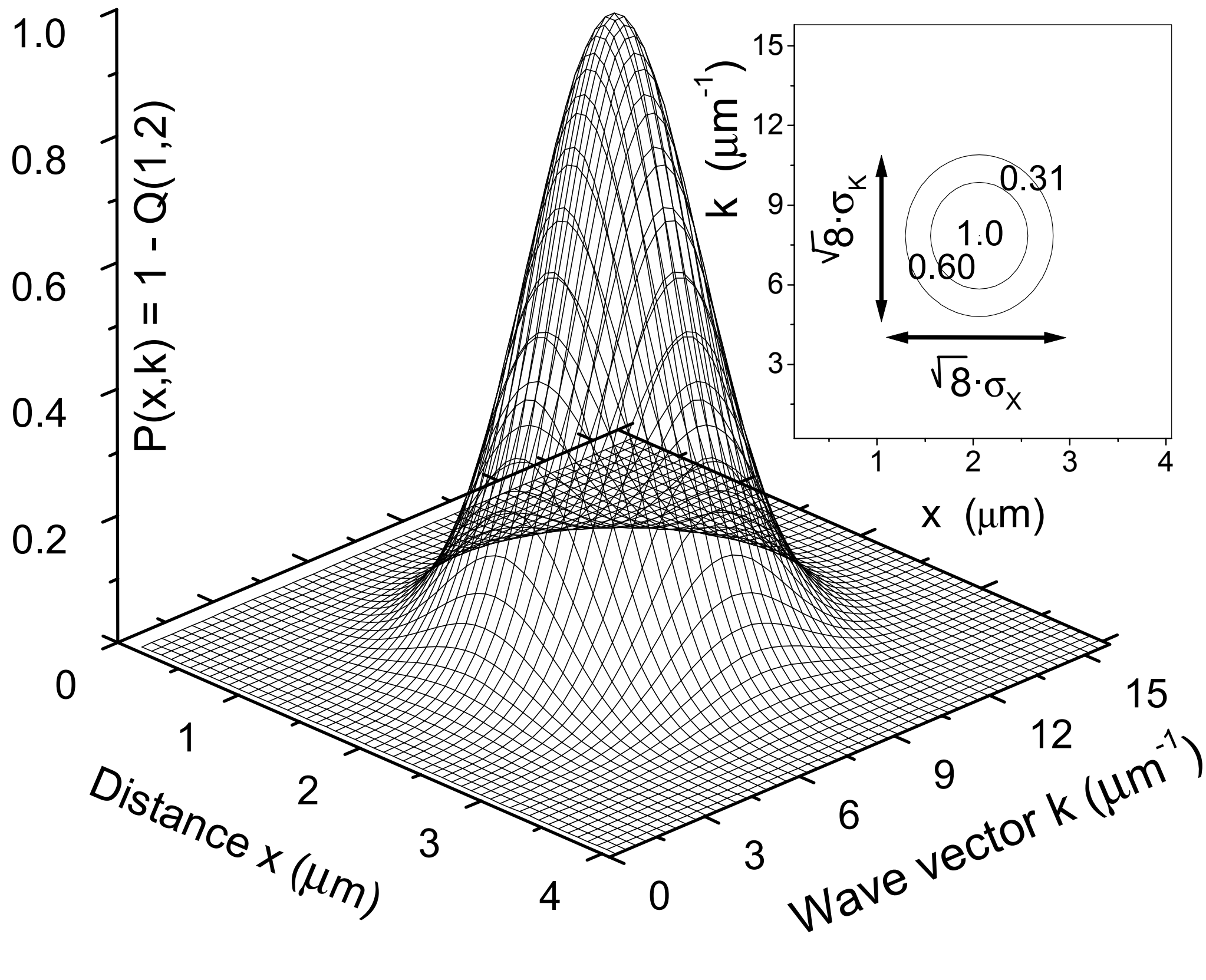

Figure 1: Xavier Oriols

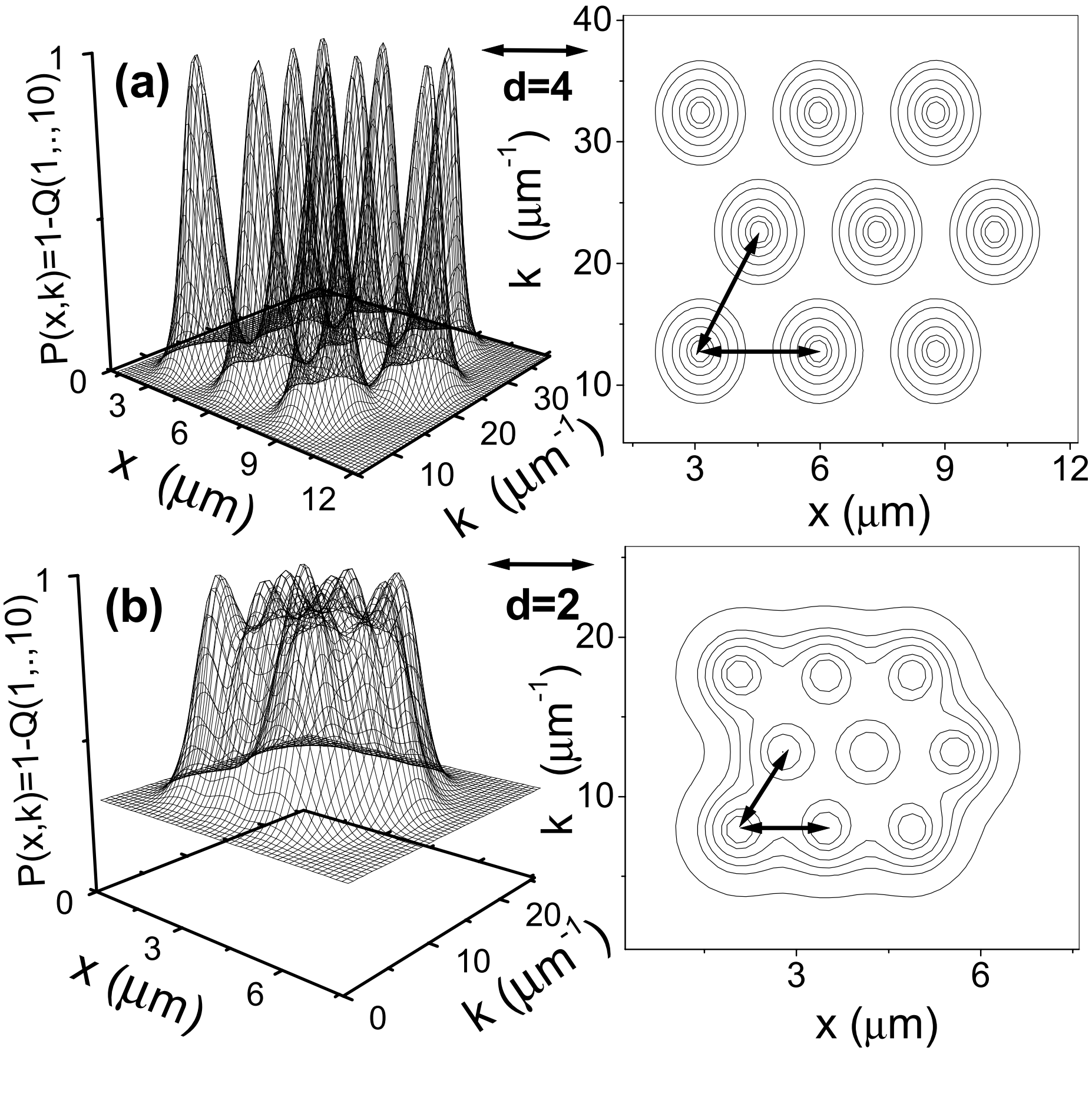

Figure 2: Xavier Oriols

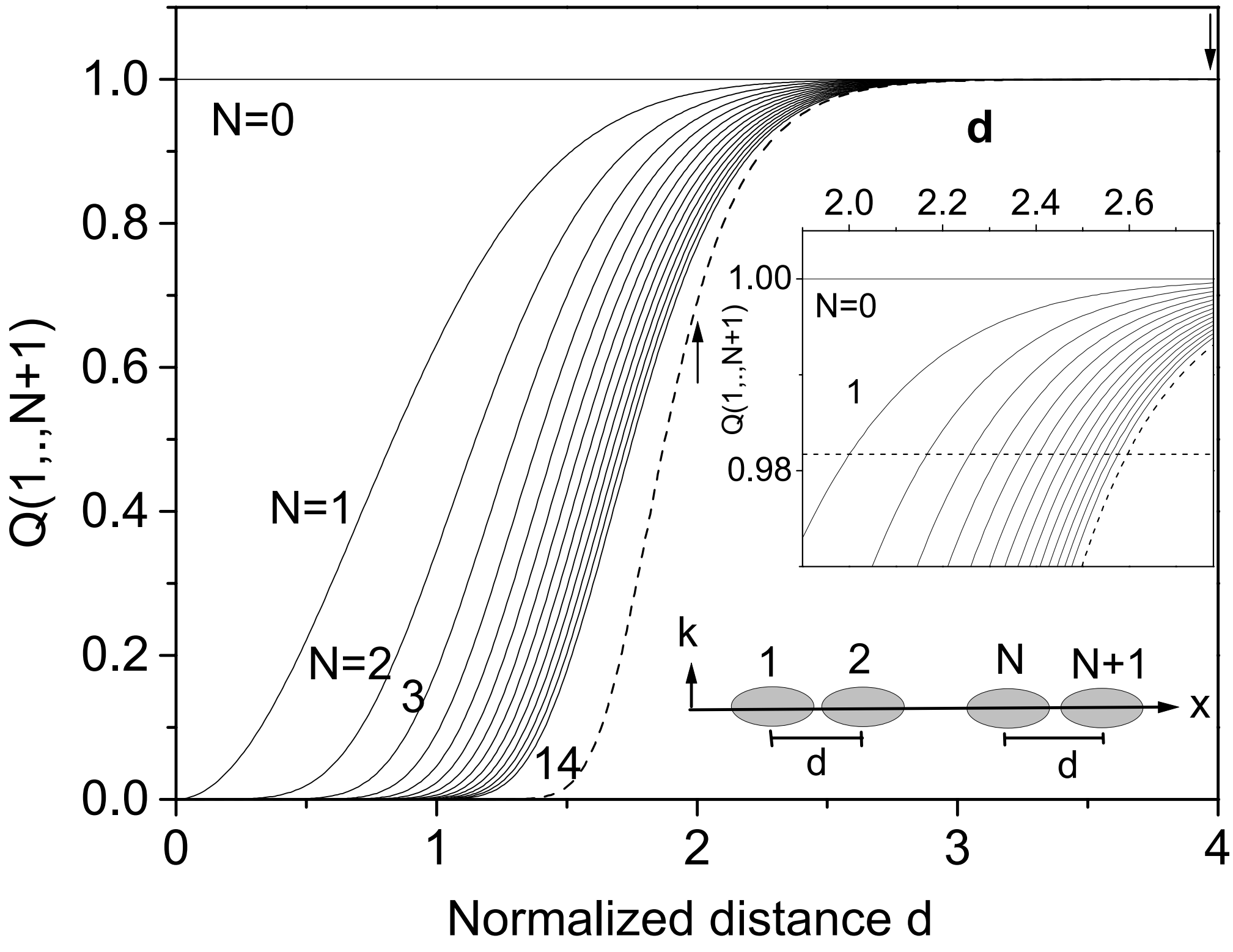

Figure 3: Xavier Oriols

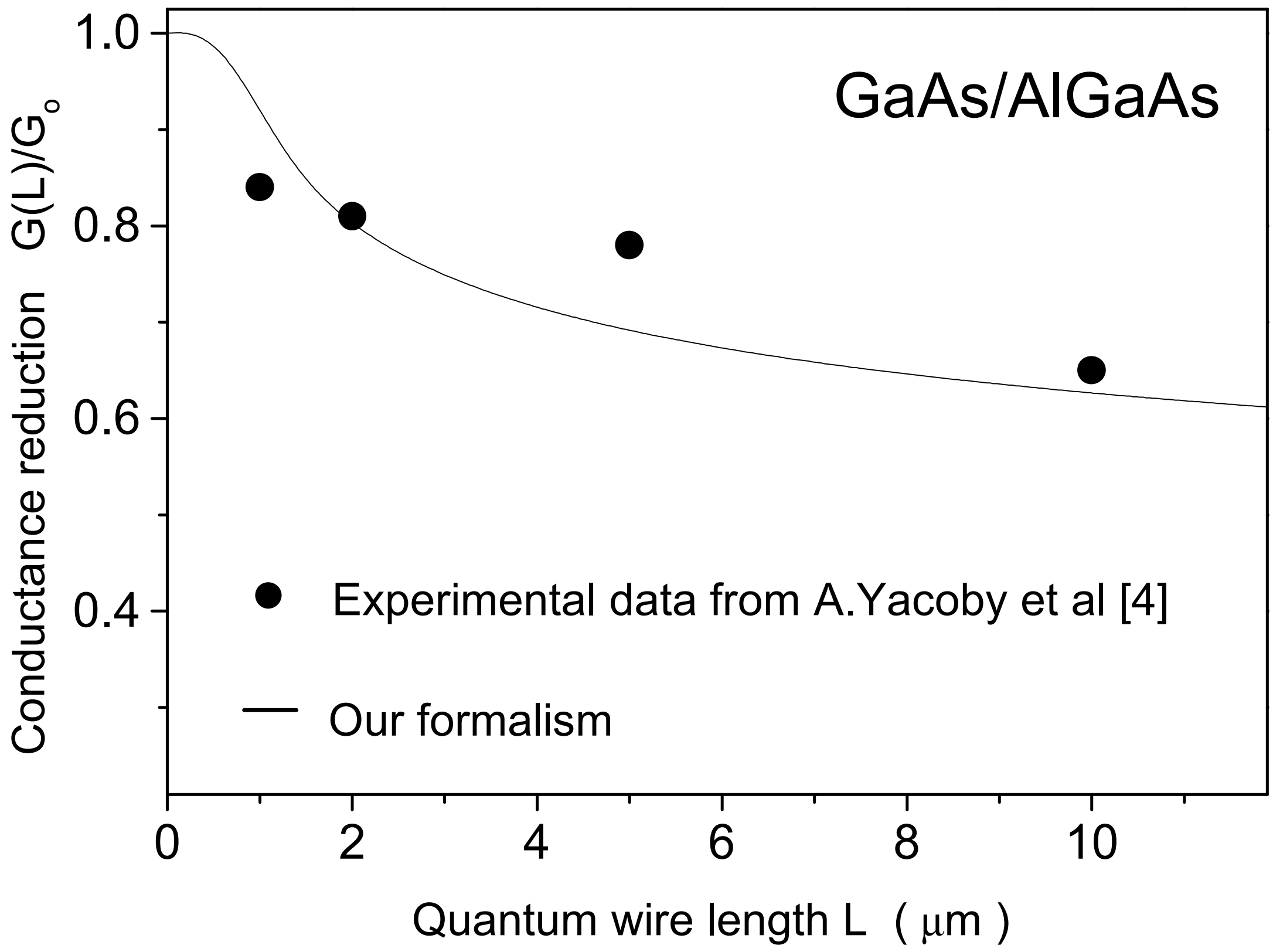

Figure 4: Xavier Oriols